\documentclass[pra,aps,twocolumn,10pt,showpacs,groupedaddress,superscriptaddress,floatfix]{revtex4-1}
\usepackage{amsmath,amsfonts,amssymb,graphics,graphicx,epsfig,color,times}
\usepackage[utf8]{inputenc}
\usepackage{color}
\usepackage{bbm, dsfont} 
\usepackage{subfigure}
\usepackage{hyperref}
\usepackage{mathrsfs}
\usepackage{verbatim}
\begin{document}
\title{Directional subradiance from helical-phase-imprinted multiphoton states}
\author{H. H. Jen}
\email{sappyjen@gmail.com}
\affiliation{Institute of Physics, Academia Sinica, Taipei 11529, Taiwan}

\date{\today}
\renewcommand{\k}{\mathbf{k}}
\renewcommand{\r}{\mathbf{r}}
\newcommand{\parallelsum}{\mathbin{\!/\mkern-5mu/\!}}
\def\p{\mathbf{p}}
\def\R{\mathbf{R}}
\def\bea{\begin{eqnarray}}
\def\eea{\end{eqnarray}}
\def\ba{\begin{array}}
\def\ea{\end{array}}
\def\bdm{\begin{displaymath}}
\def\edm{\end{displaymath}}
\def\red{\color{red}}
\begin{abstract}
We theoretically investigate the far-field scattering properties of multiphoton super- and subradiant states which can be prepared by multiphoton excitations with orbital angular momentum (OAM).\ Due to multiphoton interference, the far-field patterns of the subradiant modes show directional scattering along the excitation direction or transverse scattering with number of peaks equal to the number of atoms.\ When more atoms are involved, we consider structures of stacked and concentric rings, which respectively show enhanced directional scattering and smoothed emission patterns.\ Our scheme gives insights to prepare many-body subradiant states, and is potentially applicable to quantum storage of multiphoton with OAM.\ By designing atomic spatial distributions, these cooperative states can tailor the far-field emission properties, which is useful for light collections and quantum information manipulations. 
\end{abstract}
\maketitle
\section{Introduction}
Superradiance \cite{Dicke1954, Gross1982} results from cooperative light-matter interactions, which typically radiates directionally \cite{Rehler1971} with a shortened time scale.\ This enhancement of the spontaneous emission comes from the resonant dipole-dipole interactions (RDDI) \cite{Lehmberg1970} induced by rescattered photons between atoms.\ In such strong coupling regime, subradiance on the other hand also becomes evident \cite{Guerin2016} in the afterglow \cite{Mazets2007} of the superradiance.\ The remnant radiation with slow decay rates signifies the presence of subradiant states, and at the same time, indicates that the initially excited superradiant state is not the eigenstate of the dynamically interacting system.\ This is due to the long-range nature of RDDI which couples every other atom in the system, and thus in general all Hilbert space of super- and subradiant states are involved in the dissipation process.\ This offers an opportunity to manipulate light-matter interacting systems \cite{Kimble2008, Hammerer2010} and to deterministically access these subradiant states, for example, in one-dimensional \cite{Jen2016_SR} or three-dimensional (3D) atomic arrays \cite{Jen2016_SR2, Jen2017_MP}. 

In addition to the obvious linewidth broadening (superradiance) or narrowing (subradiance) in fluorescence experiments \cite{Guerin2016}, collective light scattering in a dense atomic medium \cite{Bromley2016,Zhu2016,Shahmoon2017} also provides useful information of cooperative light-matter interacting systems.\ Aside from well-known directional light scattering of superradiance, recently we propose to use helical-phase-imprinted (HPI) single-photon subradiant states \cite{Jen2018_far} on a ring array to investigate the far-field emission patterns.\ This subradiant state can be prepared by absorption of single-photon carrying the orbital angular momentum (OAM) \cite{Arnaut2000, Mair2001, Molina2007, Dada2011, Fickler2012} and becomes the alternative candidate for storage of light quanta with OAM \cite{Nicolas2014, Ding2015, Zhou2015}.\ Similar ring-shaped lattice has been proposed to excite high-lying Rydberg states \cite{Olmos2010} which lead to nonclassical collective photon emissions.\ 

In this paper, we investigate the far-field scattering properties of HPI multiphoton subradiant states in the ring arrays, which are less studied systematically.\ The HPI multiphoton states can tailor and modify the radiation properties by controlling and engineering the atomic configurations and light polarizations.\ Thus, this is potentially useful in generating and detecting entangled multiphoton states \cite{Pan2012}, and also offers opportunities in manipulating multiphoton resource in quantum network \cite{Kimble2008}.\ In Sec. II, we construct the HPI multiphoton states which can be created when a multiphoton source with OAM is absorbed by the atomic ring arrays.\ We then derive the general expression of far-field scattering out of these multiphoton states.\ In Sec. III, we use stacked and concentric ring structures to investigate their far-field emission properties.\ In Sec. IV, we discuss potential applications in quantum state engineering, and give a conclusion.

\section{Helical-phase-imprinted multiphoton states}

When a resonant multiphoton source of $M$ photons is absorbed by $N$ two-level atoms ($|g\rangle$ and $|e\rangle$), a symmetric state can be formed, and such that $M$ excited and $N$ $-$ $M$ ground states are symmetrically distributed in all atoms with equal probabilities $(C^N_M)^{-1}$ where $C$ denotes the binomial coefficients.\ To access the other possible multiphoton super- and subradiant states, we have proposed to use phase-imprinting methods to unitarily transform the system into \cite{Jen2017_MP}, 
\bea
\left|\phi_n^{(M)}\right\rangle&&=\prod_{j=1}^{M}\sum_{\mu_j=\mu_{j-1}+1}^{N-M+j}\frac{e^{i\k_L\cdot\mathbf{R}_M}}{\sqrt{C^N_M}}e^{i\frac{2n\pi}{C^N_M}[f(\vec{\mu})-1]}
\left|\psi_N^{(M)}(\vec{\mu})\right\rangle,\label{multiphoton}\nonumber\\
\eea
where $n$ $\in$ [$1$, $C^N_M$], and $f(\vec{\mu})$ $=$ $\sum_{i=1}^M\mu_i$ indicates the linear increases of the imprinted phases in a 3D atomic array.\ We use a concise representation of the traveling phase $\k_L\cdot\mathbf{R}_M$ $\equiv$ $\k_L\cdot\sum_{m=1}^M \r_{\mu_m}$, and define the bare states with $M$ excitations, $|\psi_N^{(M)}(\vec{\mu})\rangle$ $\equiv$ $|e\rangle_{\mu_1}|e\rangle_{\mu_2}$ $\cdots$ $|e\rangle_{\mu_M}|g\rangle^{\otimes(N-M)}$ with $\vec{\mu}$ $=$ $(\mu_1,$ $\mu_2,$ $\cdots,\mu_M)$.\ Note that this specific construction of Hilbert space is not unique, and there are many ways to create such multiple excitations space.\ This phase imprinting method dynamically controls the incrementally increasing phases either by Zeeman field or a Stark field gradient pulse, but demands large field gradient strengths or long interaction time.\ 

\begin{figure}[t]
\centering
\includegraphics[width=8.0cm,height=5.5cm]{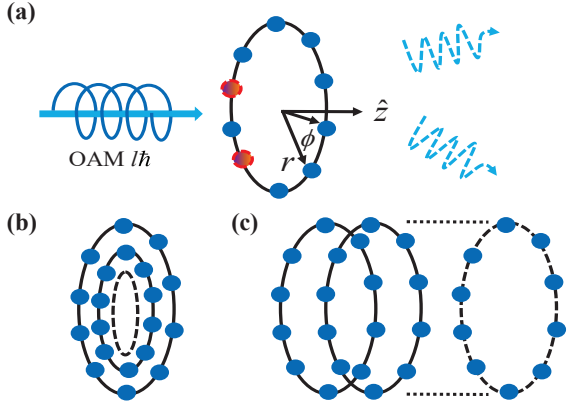}
\caption{Schematic helical-phase-imprinted multiphoton state preparation.\ (a) A multiphoton source carrying orbital angular momentum (OAM) is absorbed by the atoms on a single ring, where two excitations (dash circles) are shown here for demonstration and as one possibility out of the complete excitation space.\ The atomic system is then prepared into one of multiphoton super- or subradiant states, depending on OAM $l\hbar$ of light.\ The scattered light property out of these multiphoton states depends on the ring geometry of radius $r$ and $\phi$.\ Other atomic configurations of (b) concentric or (c) stacked rings can be the candidates for creation of helical-phase-imprinted multiphoton states.}\label{fig1}
\end{figure}

Nevertheless, this linearly increasing phase can be also imprinted in an azimuthal direction by light with OAM \cite{Jen2018_far} without external field gradients.\ We consider Laguerre-Gaussian modes of light excitations, which carries a phase dependence of $e^{il\phi}$ with a quantized OAM of $l\hbar$.\ For $N_\phi$ equidistant atoms sitting on a single ring in Fig. \ref{fig1}, the light imprints the phase of $\phi$ $=$ $2\pi(\sum_{m=1}^M\mu_m-1)/N_\phi$ on the atoms for $M$ excitations.\ To label the ring structures of concentric or stacked rings in Figs. \ref{fig1}(b) and (c), we raise up atomic indices to superscripts and denote $\mu^j$ $=$ $(\mu_r^j-1)N_\phi$ $+$ $\mu_\phi^j$ or $(\mu_z^j-1)N_\phi$ $+$ $\mu_\phi^j$, respectively.\ The number of integrated rings is $\mu_{r,z}^j$ $\in$ $[1,N_{r,z}]$ where $N_{r,z}$ represents the rings along $\hat r(z)$ direction, and the total number of atoms becomes $N$ $=$ $N_{r,z}N_\phi$.\ For each atom labeled as $j$, $\mu_\phi^j$ is within $[1,N_\phi]$.\ Therefore, the light with OAM absorbed by the ring array forms HPI multiphoton states
\bea
\left|\phi_l^{(M)}\right\rangle_{\rm HPI}=&&\prod_{j=1}^{M}\sum_{\mu^j=\mu^{j-1}+1}^{N-M+j}\frac{e^{i\k_L\cdot\mathbf{R}_M^j}}{\sqrt{C^N_M}}\nonumber\\
&&\times e^{i\frac{2l\pi}{N_\phi}[f^j(\vec{\mu}_\phi)-1]}
\left|\psi_N^{(M)}(\vec{\mu})\right\rangle,\label{HPIMP}
\eea
where here we note that $\vec{\mu}_\phi$ $=$ $(\mu^1_\phi,$ $\mu^2_\phi,$ $\cdots,\mu^M_\phi)$, and define $\R_M^{j}$ $\equiv$ $\sum_{j=1}^Mr_{\mu^{j}}$, $f^{j}(\vec\mu_\phi)$ $\equiv$ $\sum_{j=1}^M\mu_\phi^{j}$.

Comparing Eq. (\ref{multiphoton}), the HPI multiphoton states of Eq. (\ref{HPIMP}) has restricted imprinted phases by OAM of light since it puts phases along $\hat \phi$ direction only.\ Thus, Eq. (\ref{HPIMP}) forms only a subspace of multiply-excited states of Eq. (\ref{multiphoton}).\ However, HPI multiphoton states can expedite the preparation protocols by simply using light with OAM, which removes the requirement of external addressing fields and makes the proposed scheme here more efficient and dominant in generating multiphoton subradiant states.\ In the next subsection, we further derive the general expression for far-field emission properties of HPI multiphoton states, which we use to investigate the effect of multiphoton interference and atomic configurations on the emission patterns in the next section.    

\subsection{Far-field scattering}

The far-field scattering patterns can provide a unique characterization of the HPI multiphoton states.\ The general expression of the far-field scattering from uniformly-polarized two-level atoms is \cite{Lehmberg1970}
\bea
\left\langle\vec E^*(\R,t')\vec E(\R,t)\right\rangle=&&\left(\frac{\omega_{eg}^2}{4\pi\epsilon_0 c^2}\right)^2\frac{|\vec p|^2}{R^2}\left[1-(\hat \R\cdot \hat \p)^2\right]\nonumber\\
&&\times\sum_{\alpha,\beta=1}^N e^{i\k_R\cdot\r_{\alpha\beta}}\left\langle\hat{\sigma}_\alpha^+(t')\hat{\sigma}_\beta^-(t)\right\rangle,\nonumber\\\label{intensity}
\eea
where $\omega_{eg}$ is the energy difference, $\hat \R$ $\parallelsum$ $\k_R$ denotes the observation direction, $\r_{\alpha\beta}$ $=$ $\r_\alpha$ $-$ $\r_\beta$ denotes mutual atomic separations, and the dipole orientation $\hat \p$ is determined by the polarization of excitation field.\ Since the far-field scattering concerns the observation point way distant from the atoms, the radiation phase is then contributed by the retarded ones $e^{-i\k_R\cdot\r_\alpha}$ and $|\k_R|$ $=$ $\omega_{eg}/c$.\ At equal time of Eq. (\ref{intensity}), we obtain the far-field radiation patterns from the HPI multiphoton states $|\phi_l^{(M)}(t)\rangle_{\rm HPI}$,
\bea
\Omega_f(\theta,\phi)\equiv&&\frac{\left\langle\vec E^*(\R,t)\vec E(\R,t)\right\rangle}{I_0(t)},\nonumber\\
=&&\left[1-(\hat \R\cdot \hat \p)^2\right]
\prod_{m=1}^{M}\sum_{\mu^m=\mu^{m-1}+1}^{N-M+m}\prod_{j=1}^{M}\sum_{\mu^j=\mu^{j-1}+1}^{N-M+j}\nonumber\\
&&\times \frac{e^{i(\k_R-\k_L)\cdot(\R_M^{m}-\R_M^{j})}}{C^N_M}e^{\frac{i2l\pi}{N_\phi}[f^j(\vec\mu_\phi)-f^m(\vec\mu_\phi)]},\label{far}
\eea
where $I_0(t)$ is the time-evolving fluorescence intensity and depends on the eigenvalues of the coupling matrix \cite{Jen2016_SR, Jen2016_SR2, Jen2017_MP}.\ An implicit dependence of multiply-excited configurations is in the spatial and imprinted phases respectively.\ Equation. (\ref{far}) characterizes the far-field scattering property from the HPI multiphoton states created by an excitation field with $l\hbar$ OAM, which involves the interplay of the atomic distributions $\r_{\mu^j}$ and the helical phases $\propto$ $e^{i2l\pi/N_\phi}$.\ In the next section, we investigate the far-field scattering properties for the atomic configurations in Fig. \ref{fig1} and various OAM of light.

\section{Light scattering from HPI multiphoton states}

Here we focus on HPI two- and three-photon states prepared in a single, stacked, and concentric rings.\ For $M$ excitations, there are in general $C^N_M$ possible HPI multiphoton states.\ Before we investigate the case of many atoms, a few atoms study can illuminate the interplay between imprinted phases, atomic spatial distributions, and multiphoton interference.

\subsection{Three- and four-atom case}

For three atoms on a ring with the same arc length, we have the far-field scattering from HPI two-photon states from $\hat x$-polarized light,
\bea
\Omega_f(\theta,\phi)=&&(1-\sin^2\theta\cos^2\phi)\left[3+\left(e^{i\sqrt{3}\bar r\sin\theta\sin\phi+i\frac{2l\pi}{3}}\right.\right.\nonumber\\
&&+e^{i\frac{3}{2}\bar r\sin\theta\cos\phi+i\frac{\sqrt{3}}{2}\bar r\sin\theta\sin\phi+i\frac{4l\pi}{3}}\nonumber\\
&&\left.\left.+e^{i\frac{3}{2}\bar r\sin\theta\cos\phi-i\frac{\sqrt{3}}{2}\bar r\sin\theta\sin\phi+i\frac{2l\pi}{3}}+\textrm{c.c}\right)\right],\label{three}
\eea
where $\bar r$ $=$ $|\k_R|r$ and c.c. means the complex conjugate of counterparts in the parenthesis.\ This far-field pattern is exactly the same as HPI single photon states, since the traveling phases contributed from atomic configurations in Eq. (\ref{far}), $\R_M^m$ $-$ $\R_M^j$, involve only pairwise atoms.\ This is also the case for the difference of imprinted helical phases, $f^{j(m)}(\vec\mu_\phi)$.\ Therefore, the far-field patterns of HPI multiphoton states with $M$ $=$ $N-1$ exactly correspond to HPI single photon states on a ring \cite{Jen2018_far}.

\begin{figure}[b]
\centering
\includegraphics[width=8.0cm,height=4.5cm]{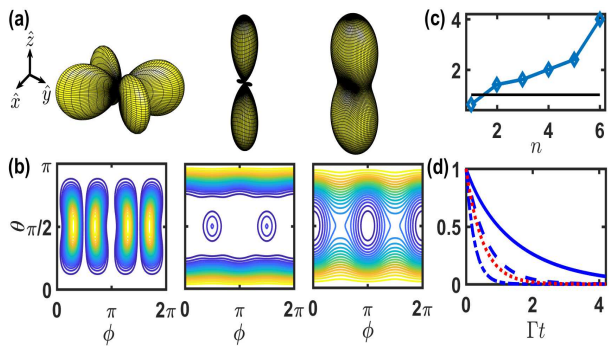}
\caption{The far-field $\Omega_f(\theta,\phi)$ of $\hat x$-polarized HPI two-photon states for four atoms on a ring.\ (a) Three-dimensional plots of $\Omega_f(\theta,\phi)$ at $r/\lambda$ $=$ $0.2$ with $l$ $=$ $1$, $2$, and $4$, respectively.\ Corresponding two-dimensional plots are shown in (b).\ (c) The spontaneous decay rates, Re$[-\lambda_n]/(\Gamma/2)$ in an ascending order, are numerically calculated from the eigenvalues $\lambda_n$ for this given atomic configuration.\ A horizontal line guides the eye for natural decay rate Re$[-\lambda_n]=\Gamma/2$.\ (d) The fluorescence intensity $I_0$(t) of the HPI two-photon states with $l$ $=$ $1$ (dash), $2$ (solid), and $4$ (dash-dot), compares with $e^{-2\Gamma t}$ (dot).}\label{fig2}
\end{figure}

The unique emission pattern of HPI multiphoton states then emerges in two-photon states of a ring with four atoms.\ In Fig. \ref{fig2}, we show the scattering patterns $\Omega_f(\theta,\phi)$ with associated fluorescence intensity $I_0(t)$.\ Since the HPI multiphoton states are not the eigenstates of the system in general \cite{Jen2017_MP}, their fluorescence results from several eigen decay rates which we show in Fig. \ref{fig2}(c) for the specific atomic configuration with six ($C^N_M$) eigenstates.\ We then distinguish the sub- ($l$ $=$ $1$, $2$) from superradiant ($l$ $=$ $4$) states by comparing $I_0(t)$ with the intrinsic decay function $e^{-2\Gamma t}$.\ A clear signature of the subradiant far-field pattern is the side scattering \cite{Bromley2016, Zhu2016, Jen2018_far} in the mode of $l$ $=$ $1$, while the superradiant mode of $l$ $=$ $4$ typically shows a forward and backward scattering along the excitation direction $\hat z$.

Interestingly, the most subradiant state of $l$ $=$ $2$ shows not only small side scattering peaks at $\phi$ $=$ $\pi/2$ and $3\pi/2$, but also enhanced forward and backward scatterings, bearing both signatures of single-photon sub- and superradiance.\ This is in contrast to HPI single-photon subradiant states \cite{Jen2018_far} which allow only side scatterings.\ In addition, this illustrates the multiphoton interference in the HPI states, which carries an extra degree of freedom in phase-imprinting other than atomic configurations (traveling phases) and excitation modes (various $l$).\ We note that discrete $C_4$ rotational symmetry in $\Omega_f(\theta,\phi)$ is preserved for linear polarizations in the atomic system of $N$ $=$ $4n$ with integers $n$.\ This can be seen as we change $\phi$ to $\phi+\pi/2$ in the prefactor of Eq. (\ref{far}), which changes $\hat x$-polarized emission pattern to $\hat y$-polarized one.\ Also $l$ $\leftrightarrow$ $-l$ symmetry in $\Omega_f(\theta,\phi)$ is sustained for all $N$ as $r$ $\rightarrow$ $0$.\ In general for all $r$, even $N$ preserves this mode symmetry.\ This mode symmetry emerges due to the regular convex polygon geometry which we assume for atoms forming a ring and incremental helical phases we apply, and as such the far-field scattering in Eq. (\ref{far}) does not change as $l$ $\rightarrow$ $-l$ (or equivalently $N-l$).\ Therefore, $\Omega_f(\theta,\phi)$ of $l$ $=$ $3$ is the same as $l$ $=$ $1$ in Fig. \ref{fig2}.\ For another example of $N$ $=$ $6$ with $M$ $=$ $2$, one of the far-field scattering contributions, $\cos[\k_R\cdot(\r_1+\r_2-\r_3-\r_4)+2l\pi\times 4/6]$, pairs up with $\cos[\k_R\cdot(\r_4+\r_5-\r_1-\r_6)+2l\pi\times (-2)/6]$, which interchanges to each other as $l$ $\rightarrow$ $-l$.


\subsection{Single ring} 

\begin{figure}[t]
\centering
\includegraphics[width=8.0cm,height=4.5cm]{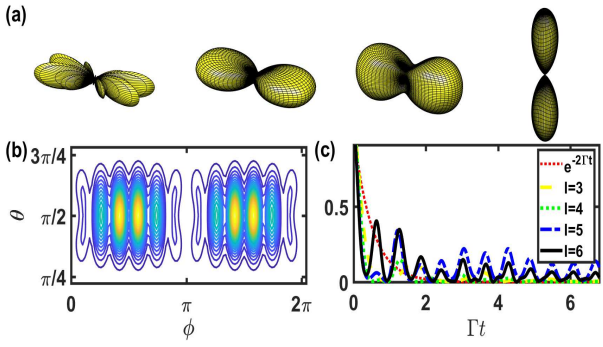}
\caption{The far-field $\Omega_f(\theta,\phi)$ of $\hat x$-polarized HPI two-photon states for $N$ $=$ $12$ on a ring.\ (a) Three-dimensional plots of $\Omega_f(\theta,\phi)$ with $l$ $=$ $3-6$ with a specific two-dimensional plot of $l$ $=$ $3$ in (b).\ (c) The fluorescence intensity $I_0$(t) of the HPI two-photon states for $l$ $=$ $3$-$6$, compares with $e^{-2\Gamma t}$.\ The viewing angle and $r$ are the same as in Fig. \ref{fig2}(a).}\label{fig3}
\end{figure}

Here We take $N$ $=$ $12$ on a ring as an example and investigate the far-field scattering patterns of HPI two-photon subradiant states.\ This $N$ has $C_4$ rotational symmetry so we use $\hat x$-polarized light without loss of generality.\ In Fig. \ref{fig3}(a), we show the side scattering of the subradiant modes with $l$ $=$ $3$, $4$, and $5$.\ The specific two-dimensional plot of $l$ $=$ $3$ mode is illustrated in Fig. \ref{fig3}(b), which reveals $N$ side scattering peaks.\ This specific mode corresponds to the number of atoms, similar for $l$ $=$ $1$ mode in Fig. \ref{fig2}(a), and thus can be used to characterize the atomic configuration.\ The $l$ $=$ $6$ subradiant mode in Fig. \ref{fig3}(a), also similar to the $l=2$ mode in Fig. \ref{fig2}(a), shows forward and backward scattering, but radiates in a rate only second to the most subradiant mode of $l$ $=$ $5$.\ This can be seen in the time evolutions of the subradiant modes in Fig. \ref{fig3}(c) where $l$ $=$ $5$ state evolves longer than the other subradiant modes since it occupies the most subradiant eigen decay rate.\ The HPI states of $l$ $=$ $3$, $4$, and $5$ drop rapidly initially due to partial overlaps with the superradiant eigenmodes while later they still evolve past the intrinsic decay $e^{-2\Gamma t}$, indicating a subradiant afterglow.\ 

\begin{figure}[b]
\centering
\includegraphics[width=8.0cm,height=4.0cm]{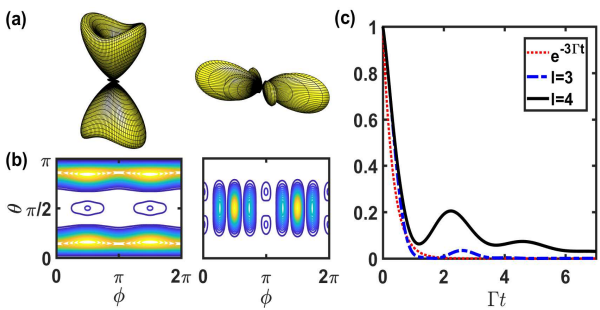}
\caption{The far-field $\Omega_f(\theta,\phi)$ of $\hat x$-polarized HPI three-photon states for $N$ $=$ $8$ on a ring.\ (a) Three-dimensional plots of $\Omega_f(\theta,\phi)$ of subradiant states with $l$ $=$ $3$ and $4$.\ Corresponding two-dimensional plots are shown in (b).\ (c) The fluorescence intensity $I_0$(t) of the HPI three-photon states with $l$ $=$ $3$ and $4$, compares with $e^{-3\Gamma t}$.\ The viewing angle and other parameters are the same as in Fig. \ref{fig2}(a).}\label{fig4}
\end{figure}

For far-field scattering patterns of HPI three-photon subradiant states, we show in Fig. \ref{fig4} with $N$ $=$ $8$.\ We note that as $N$ increases, the total Hilbert space goes up as $\mathcal{O}(N^3)$, so here we only focus on the feature of directional subradiance instead of multi-atom effect.\ For larger $N$ on the other hand, we will investigate the effect of stacked and concentric rings in the next subsection.\ In Fig. \ref{fig4}, we show the effect of multiphoton interference on the far-field scattering, where the subradiant mode of $l$ $=$ $N/2$ encodes the information of $N$ side scattering peaks, in contrast to forward/backward scattering $l=N/2$ modes in Figs. \ref{fig2} and \ref{fig3}.\ Furthermore, the subradiant mode at $l$ $=$ $3$ in Fig. \ref{fig4} shows an oblique scattering, instead of forward and backward directions.\ 

The HPI multiphoton subradiant states show quite rich far-field scattering patterns.\ Except for side scatterings, the subradiant emission also allows along or near the excitation direction.\ In addition, the oscillation of time evolutions manifests the beating frequency from the differences of cooperative frequency shifts \cite{Jen2016_SR}.\ This shows flexibility to tailor and manipulate the emission time and directions of multiphoton states, and enables the one to one correspondence between the atomic configurations and the far-field scattering patterns.\

\subsection{Stacked and concentric rings} 

\begin{figure}[t]
\centering
\includegraphics[width=8.0cm,height=4.0cm]{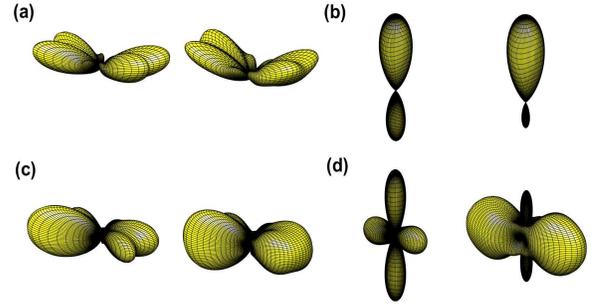}
\caption{The far-field $\Omega_f(\theta,\phi)$ of $\hat x$-polarized HPI two-photon states for $N_\phi$ $=$ $8$ and $N_{r,z}$ $=$ $2$ and $3$ in the stacked and concentric rings.\ For stacked rings, the subradiant states with (a) $l$ $=$ $2$ and (b) $l$ $=$ $4$ are illustrated with $N_{z}$ $=$ $2$ and $3$ from left to right.\ Similarly for concentric rings, the subradiant (left) and superradiant (right) states respectively with (c) $l$ $=$ $2$ and (d) $l$ $=$ $4$ are illustrated with $N_{r}$ $=$ $2$ and $3$ from left to right.\ The viewing angle and other parameters are the same as in Fig. \ref{fig2}(a), and $d_z/\lambda$ $=$ $0.35$.}\label{fig5}
\end{figure}

Finally we investigate the far-field scattering properties of HPI multiphoton subradiant states in the stacked and concentric rings.\ We have investigated these two configurations of HPI single-photon states \cite{Jen2018_far} which show enhanced forward scattering and narrowing effect on side scattering peaks in the stacked and concentric rings respectively.\ In Fig. \ref{fig5}, we show the far-field scattering properties of HPI two-photon subradiant states with $l$ $=$ $2$ and $4$, in the stacked and concentric rings respectively.\ Similar to HPI single-photon states, the forward scattering is enhanced in Figs. \ref{fig5}(a) and (b) as $N_z$ increases.\ On the contrary, comparing Fig. \ref{fig3}(a), the transverse scattering peaks are smoothed out due to multiphoton interference.\ This smoothing effect is even more significant in the concentric rings of Fig. \ref{fig5}(c) as $N_r$ increases, and the states of $l$ $=$ $2$ and $4$ become superradiant for a larger $N_r$.\ In Fig. \ref{fig5}(d), the effect of concentric rings weakens the forward-backward scattering along with emerging side scattering.\ Therefore, in the perspective of preparing multiphoton subradiant states in a large atomic system, the stacked rings are better than the concentric ones since the former allows a more directional emission pattern which is better for light collection and detection.

\begin{figure}[t]
\centering
\includegraphics[width=8.0cm,height=4.0cm]{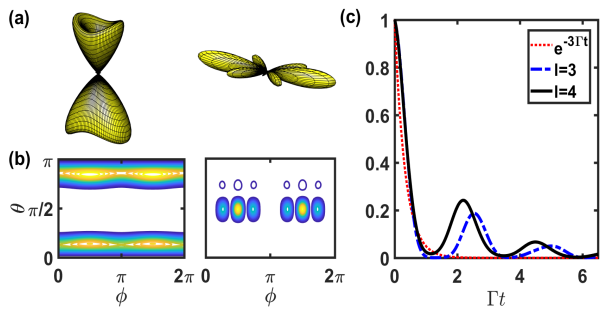}
\caption{The far-field $\Omega_f(\theta,\phi)$ of $\hat x$-polarized HPI three-photon states for $N_\phi$ $=$ $8$ and $N_{z}$ $=$ $2$ in the stacked rings.\ The subradiant states with (a) $l$ $=$ $3$ and $l$ $=$ $4$ are illustrated from left to right with corresponding two-dimensional plots in (b).\ (c) The fluorescence intensity $I_0$(t) of the HPI three-photon states compares with $e^{-3\Gamma t}$ (dot).\ The viewing angle and other parameters are the same as in Fig. \ref{fig5}.}\label{fig6}
\end{figure}

For HPI three-photon subradiant states, we show the results of $\Omega_f(\theta,\phi)$ of two stacked rings in Fig. \ref{fig6} as a comparison to Fig. \ref{fig4}.\ We find the enhanced oblique scattering toward the backward direction for $l$ $=$ $3$ mode and narrowing side scattering peaks for $l$ $=$ $4$ in polar angles $\theta$.\ The small peaks at $\theta$ $\approx$ $3\pi/4$ for $l$ $=$ $4$ is also a signature of enhanced oblique scattering which breaks the mirror symmetry of the emission patterns in Fig. \ref{fig4} with respect to $\hat x$-$\hat y$.\ We note that the clear $N_\phi$ number of transverse scattering peaks disappear due to multiphoton effect.\ The time evolutions of these states in Fig. \ref{fig6}(c) shows the subradiant fluorescence, and the mode for $l$ $=$ $3$ is more subradiant than the one in Fig. \ref{fig4}(c), which indicates the cooperative enhancement in subradiance when more atoms are stacked up.\ For more atoms, for example of $N_z=3$, we find similar enhancement in oblique scattering and narrowing polar angle distribution for $l=3$ and $4$ modes respectively.\ For concentric rings, these three-photon modes do not show significant subradiance, similar to Figs. \ref{fig5}(c) and (d).\ 

\section{Discussion and Conclusion}

Recently there is a huge advance in precisely controlling atomic separations in waveguide \cite{Goban2015} or integrated color centers \cite{Sipahigil2016} and in assembling discrete atoms by optical microtraps \cite{Barredo2016, Endres2016}, which has enabled fine tunings of light-matter interactions.\ These experimental techniques have promised an ultimate quantum control in quantum optical systems.\ With well control of atomic configurations, we can manipulate RDDI of the spontaneous emission process.\ This controls the timescale of the emission, which makes subradiant states a resource for quantum memory.\ Subradiant far-field scattering, on the other hand, provides additional information for light collection and specific characterization of atomic geometry.\ This helps prepare genuine quantum states and makes quantum state engineering more efficient.

In conclusion, we propose to prepare the multiphoton subradiant states by imprinting the helical phases from OAM of light on the atoms in the stacked and concentric rings, and investigate their far-field scattering patterns.\ The directional scattering of the helical-phase-imprinted multiphoton subradiant states allows for light collimation, and their small spontaneous decay rates are preferential to quantum storage of light quanta with orbital angular momenta.\ We find enhanced forward (oblique backward) emission for HPI two (three)-photon subradiant states which can also support transverse scattering peaks.\ Therefore, we can tailor and control the scattering properties by engineering the atoms spatially, which essentially can facilitate quantum control of the many-body subradiant states and is useful in manipulating entanglement of large orbital angular momenta \cite{Fickler2016} or operating quantum gates \cite{Babazadeh2017}.\

\section*{ACKNOWLEDGMENTS}
This work is supported by the Ministry of Science and Technology (MOST), Taiwan, under the Grant No. MOST-106-2112-M-001-005-MY3.\ H.H.J is partially supported by a grant from MOST of No. 106-2811-M-001-130, as an assistant research scholar in IOP, Academia Sinica.\ We are also grateful for the support of NCTS ECP1 (Experimental Collaboration Program) and for stimulating discussions with M.-S. Chang and Y.-C. Chen.


\begin{thebibliography}{99}
\bibitem{Dicke1954} R. H. Dicke, Phys. Rev. {\bf 93}, 99 (1954).
\bibitem{Gross1982} M. Gross and S. Haroche, Phys. Rep. {\bf 93} 301 (1982).
\bibitem{Rehler1971} N. E. Rehler and J. H Eberly, Phys. Rev A {\bf 3}, 1735 (1971).
\bibitem{Lehmberg1970} R. H. Lehmberg, Phys. Rev. A {\bf 2}, 883 (1970).
\bibitem{Guerin2016} W. Guerin, M. O. Ara\'{u}jo, and R. Kaiser, Phys. Rev. Lett. {\bf 116}, 083601 (2016).
\bibitem{Mazets2007} I. E. Mazets and G. Kurizki, J. Phys. B: At. Mol. Opt. Phys. {\bf 40}, F105 (2007).
\bibitem{Kimble2008} H. J. Kimble, Nature {\bf 453}, 1023 (2008).
\bibitem{Hammerer2010} K. Hammerer, A. S. S\o{}rensen, and E. S. Polzik, Rev. Mod. Phys. {\bf 82}, 1041 (2010).
\bibitem{Jen2016_SR} H. H. Jen, M.-S. Chang, and Y.-C. Chen, Phys. Rev. A {\bf 94}, 013803 (2016).
\bibitem{Jen2016_SR2} H. H. Jen, Ann. Phys. (N. Y.) {\bf 374}, 27 (2016).
\bibitem{Jen2017_MP} H. H. Jen, Phys. Rev. A {\bf 96}, 023814 (2017).
\bibitem{Bromley2016} S. L. Bromley, B. Zhu, M. Bishof, X. Zhang, T. Bothwell, J.
Schachenmayer, T. L. Nicholson, R. Kaiser, S. F. Yelin, M. D. Lukin {\it et al.}, Nat. Commun. 7:11039 (2016).
\bibitem{Zhu2016} B. Zhu, J. Cooper, J. Ye, and A. M. Rey, Phys. Rev. A {\bf 94}, 023612 (2016).
\bibitem{Shahmoon2017} E. Shahmoon, D. S. Wild, M. D. Lukin, and S. F. Yelin, Phys. Rev. Lett. {\bf 118}, 113601 (2017).
\bibitem{Jen2018_far} H. H. Jen, M.-S. Chang, and Y.-C. Chen, arXiv 1801.00559.
\bibitem{Arnaut2000} H. H. Arnaut and G. A. Barbosa, Phys. Rev. Lett. {\bf 85}, 286 (2000).
\bibitem{Mair2001} A. Mair, A. Vaziri, G. Weihs, and A. Zeilinger, Nature (London) {\bf 412}, 313 (2001).
\bibitem{Molina2007} G. Molina-Terriza, J. P. Torres, and L. Torner, Nat. Phys. {\bf 3}, 305 (2007).
\bibitem{Dada2011} A. C. Dada, J. Leach, G. S. Buller, M. J. Padgett, and E. Andersson, Nat. Phys. {\bf 7}, 677 (2011).
\bibitem{Fickler2012} R. Fickler, R. Lapkiewicz, W. N. Plick, M. Krenn, C. Schaeff, S. Ramelow, A. Zeilinger, Science {\bf 338}, 640 (2012).
\bibitem{Nicolas2014} A. Nicolas, L. Veissier, L. Giner, E. Giacobino, D. Maxein, and J. Laurat, Nat. Photonics {\bf 8}, 234 (2014).
\bibitem{Ding2015} D.-S. Ding, W. Zhang, Z.-Y. Zhou, S. Shi, G.-Y. Xiang, X.-S. Wang, Y.-K. Jiang, B.-S. Shi, and G.-C. Guo, Phys. Rev. Lett. {\bf 114}, 050502 (2015).
\bibitem{Zhou2015} Z.-Q. Zhou, Y.-L. Hua, X. Liu, G. Chen, J.-S. Xu, Y.-J. Han, C.-F. Li, and G.-C. Guo, Phys. Rev. Lett. {\bf 115}, 070502 (2015).
\bibitem{Olmos2010} B. Olmos and I. Lesanovsky, Phys. Rev. A {\bf 82}, 063404 (2010).
\bibitem{Pan2012} J.-W. Pan, Z.-B. Chen, C.-Y. Lu, H. Weinfurter, A. Zellinger, and M. $\dot{\rm Z}$ukowsky, Rev. Mod. Phys. {\bf 84}, 777 (2012).
\bibitem{Goban2015} A. Goban, C.-L. Hung, J. D. Hood, S.-P. Yu, J. A. Muniz, O. Painter, and H. J. Kimble, Phys. Rev. Lett. {\bf 115}, 063601 (2015).
\bibitem{Sipahigil2016} A. Sipahigil, R. E. Evans, D. D. Sukachev, M. J. Burek, J. Borregaard, M. K. Bhaskar, C. T. Nguyen, J. L. Pacheco, H. A. Atikian, C. Meuwly {\it et al.}, Science {\bf 354}, 847 (2016).
\bibitem{Barredo2016} D. Barredo, S. de L\'{e}s\'{e}leuc, V. Lienhard, T. Lahaye, and A. Browaeys, Science {\bf 354}, 1021 (2016).
\bibitem{Endres2016} M. Endres, H. Bernien, A. Keesling, H. Levine, E. R. Anschuetz, A. Krajenbrink, C. Senko, V. Vuletic, M. Greiner, and M. D. Lukin, Science {\bf 354}, 1024 (2016).
\bibitem{Fickler2016} R. Fickler, G. Campbell, B. Buchler, P. K. Lam, and A. Zeilinger, PNAS {\bf 113}, 13642 (2016).
\bibitem{Babazadeh2017} A. Babazadeh, {\it et al.}, Phys. Rev. Lett. {\bf 119}, 180510 (2017).
\end{thebibliography}
\end{document}